\documentclass[12pt]{article}
\usepackage{epsfig}
\usepackage{epsfig}
\begin{document}
\def\l{{(\lambda)}}
\def\tr{{\rm tr}\, }
\def\Tr{{\rm Tr}\, }
\def\hTr{\hat{\rm T}{\rm r}\, }
\def\be{\begin{eqnarray}}
\def\ee{\end{eqnarray}}
\def\ctt{\chi_{\tau\tau}}
\def\cta{\chi_{\tau a}}
\def\ctb{\chi_{\tau b}}
\def\cab{\chi_{ab}}
\def\cba{\chi_{ba}}
\def\ptt{\phi_{\tau\tau}}
\def\pta{\phi_{\tau a}}
\def\ptb{\phi_{\tau b}}
\def\>{\rangle}
\def\<{\langle}
\def\d{\hbox{d}}
\def\pab{\phi_{ab}}
\def\lb{\label}\def\appendix{{\newpage\section*{Appendix}}\let\appendix\section%
        {\setcounter{section}{0}
        \gdef\thesection{\Alph{section}}}\section}
\renewcommand{\figurename}{Fig.}
\renewcommand\theequation{\thesection.\arabic{equation}}
\renewcommand{\thefootnote}{\fnsymbol{footnote}}
\hfill{\tt IUB-TH-051}\\\mbox{}
\hfill {\tt hep-th/0501053}\\
\vskip0.3truecm
\begin{center}
\vskip 2truecm {\Large\bf Restoring Unitarity in BTZ Black Hole}
\vskip 1.5truecm
{\large\bf
Sergey N.~Solodukhin\footnote{
{\tt s.solodukhin@iu-bremen.de}}
}\\
\vskip 0.6truecm
\it{School of Engineering and Science, \\
International University Bremen, \\
P.O. Box 750561,
Bremen 28759,
Germany}
\end{center}
\vskip 1cm
\begin{abstract}
\noindent
Whether or not system is unitary can be seen from the way it, if
perturbed,
relaxes back to equilibrium. The relaxation of semiclassical black hole
can be described in terms of correlation function which
exponentially decays with time. In the momentum space
it is represented by infinite set of complex poles to be identified
with the quasi-normal
modes. This behavior is in sharp contrast to the relaxation in unitary
theory in finite
volume: correlation function of the perturbation in this case
is quasi-periodic function of  time and, in general, is expected to
show the Poincar\'e
recurrences. In this paper we demonstrate how restore unitarity in the
BTZ black hole,
the simplest example of eternal black hole in finite volume.
We start with reviewing the relaxation in the semiclassical BTZ black
hole
and how this relaxation is mirrored in the boundary conformal field
theory
as suggested by the AdS/CFT correspondence. We analyze the sum over
$SL(2,{\bf  Z})$ images of the BTZ space-time and suggest that it does
not produce
a quasi-periodic relaxation, as one might have hoped, but results in
correlation function which decays by power law. We develop our earlier
suggestion and consider a
non-semiclassical deformation of the BTZ space-time that has  structure
of wormhole
connecting two asymptotic regions semiclassically separated by horizon.
The small deformation parameter $\lambda$ is supposed to have
non-perturbative
origin to capture the finite N behavior of the boundary theory.
The discrete spectrum of perturbation in the  modified space-time is
computed and is shown to
determine the expected  unitary behavior: the corresponding time
evolution
is quasi-periodic with hierarchy of large time scales $\ln 1/\lambda$
and
$1/\lambda$ interpreted  respectively as the Heisenberg and Poincar\'e
time scales in the system.
\end{abstract}
\vskip 1cm
\newpage

\section{Introduction}
\setcounter{equation}0
That unitarity is violated in black hole was suggested by Hawking
\cite{Hawking} following his remarkable discovery that semiclassical
black holes thermally radiate. This conclusion
has been debated (see \cite{tHooft1}, \cite{Susskind}) ever since.
In spite of considerable effort made to resolve this issue
the black hole unitarity  remains one of the most intriguing problems
in physics. String theory offers a unifying approach to the short
distance
phenomena including the gravitational interaction that is based on
intrinsically unitary
quantum mechanical picture. This explains the long-standing belief that
the
unitarity problem can be resolved in string theory. One of the most
recent
and most promising suggestions in this direction is the AdS/CFT
correspondence (see \cite{Aharony:1999ti} for review and
\cite{deBoer:2004yu} for
more recent developments) which suggests that
in certain limit string theory (or, semiclassically, supergravity
theory)
on asymptotically AdS space-time can be equivalently described in terms
of
conformal field theory (CFT) living on the boundary of the space-time.
The boundary CFT is supposed to be unitary and thus the correspondence
should provide us with complete unitary description of black holes
semiclassically appearing as solutions to the supergravity theory.

Recently, Maldacena \cite{Maldacena:2001kr} made a proposal
that unitarity restores if one takes into account the topological
diversity
of gravitational instantons approaching asymptotically same boundary
manifold.
This proposal was studied in \cite{deSitter}
in the case of an analogous problem in de Sitter space.
It was further discussed in \cite{BSS}, \cite{SS}, \cite{Barbon},
\cite{Sachs}, \cite{Siopsis:2004gj}, \cite{Porrati}. It was realized
that the
problem can be analyzed by studying  the relaxation in  black
hole and boundary theory after small perturbation (in the context of
the AdS/CFT correspondence
the relaxation was first discussed in \cite{HH}). The appropriate
quantity to look at is the
correlation function of the perturbation taken at two different moments
of
time: when the perturbation was first applied and at a later time.
Two  types of relaxation were identified \cite{BSS}:  oscillatory (or,
in
general, quasi-periodic) and
the exponential decay.
In unitary theory in finite volume the correlation
function is quasi-periodic  and, in general, is expected
to show the so-called Poincar\'e recurrences, i.e. the initial value
is approached arbitrarily close in finite (although very large) period
of time.
In the case of black hole the correlation function, however, is
exponentially
decaying so that the initial configuration is never repeated.
This temporal behavior is governed by infinite set of complex
quasi-normal modes \cite{HH}, \cite{BSS}. This is a clear manifestation
of
non-unitary nature of semiclassical black hole.
We stress that it is
the presence of horizon that makes the spectrum
of perturbation continuous \cite{BSS}, \cite{Barbon} and  shifts the
poles
to the complex region.
In the case of unitary theory in finite volume the spectrum is discrete
and
complex frequencies never appear.
Note, however, that finding a way to assign a discrete spectrum to
black hole
is not the end of the story. The spectrum should be sufficiently
non-trivial
to generate the complex time evolution: black hole is supposed to have
very large
Poincar\'e recurrence time estimated as exponential in the entropy and
also show
exponential decay for intermediate time scales during which the usual
thermodynamic description would be accurate. Thus, say, simple
equidistant spectrum
(like the one for a particle in a circle) would not work: the
recurrence time
would be too short and the time evolution be strictly periodic like
that of a clock.

In this paper we study the eternal BTZ black hole \cite{BTZ} in
three-dimensional AdS
space. We start with reviewing the relaxation in the black hole and in
the
thermal AdS and how it is mirrored in the boundary CFT \cite{BSS}.
These two spaces are only two (dominating semiclassically) members  of
much larger family of gravitational
instantons approaching asymptotically  two-dimensional torus.
The whole family includes the $SL(2, {\bf Z})$ images of the BTZ black
hole.
These are that topologically distinct geometries which we are
instructed to
sum over in the AdS/CFT correspondence.
We  analyze whether the sum over these images may actually produce a
quasi-periodic result. We conclude that even though one sums over
exponentially
decaying individual terms the sum decays much slower than exponent,
namely it
decays by power law.
The resulting correlation function is  however not quasi-periodic. It
is in
agreement with earlier analysis in \cite{BSS}, \cite{Barbon},
\cite{Porrati}. This suggests
that we should look for solution to the unitarity problem within a
nonsemiclassical description of black hole.

We follow our earlier suggestion \cite{SS} and consider a
nonsemiclassical deformation of the
BTZ space-time. The deformation parameter $\lambda$ is supposed to have
non-perturbative origin, $\lambda\sim e^{1/4G}$, so that on the
boundary side the deformation should
account for finite N behavior of the boundary theory. In the deformed
metric
horizon is replaced by throat that connects two
asymptotic regions separated by horizon in the semiclassical BTZ black
hole.
The complete geometry resembles that of wormhole.
Once horizon has been removed the complex quasi-normal frequencies
disappear and the
spectrum becomes discrete and real. We compute the spectrum and find
that it has
the form of the massive spectrum
with mass $m$ proportional to $\lambda$. This spectrum appears to be
universal
since it relies only on modification of the geometry in the
near-horizon region.
The spectrum naturally incorporates hierarchy of two large time scales,
namely $\ln 1/\lambda$ and
$1/\lambda$.  The second time scale is the largest in the system and
describes
the long time correlations. Moreover it has the right value  and
can be naturally identified with the black hole Poincar\'e recurrence
time.

The paper is organized as follows. In section 2 we review the
description of
relaxation in the BTZ black hole and in boundary CFT and how this fits
in the AdS/CFT correspondence. This part is  based on the paper
\cite{BSS}. In subsection 2.3.2 we analyze the sum over $SL(2,{\bf Z})$
family.  In section 3 we introduce the nonsemiclassical modification of
the BTZ
metric and explain how finite entropy originates in this
metric.  In section 4 we apply the AdS/CFT rules and calculate the
conformal
anomaly in the boundary theory. The anomaly occurs to be non-vanishing
that we
interprete as a manifestation of a (nonperturbative) mass gap in the
boundary theory.
In section 5 we compute the spectrum for scalar perturbation in the
background
of the modified BTZ metric and show that it determines the
quasi-periodic
time evolution with two large time scales interpreted as the Heisenberg
and
Poincar\'e time scales. We conclude in section 6.

\bigskip

\bigskip

\section{Relaxation and the unitarity problem}
\setcounter{equation}0
The way how  thermal system reacts on small pertubation and whether the
subsequent time evolution drives the system back to thermal equilibrium
gives us important information about the system and the nature of its
thermal
state. Two possible types of reaction are known: the perturbation may
exponentially
decay so that the system relaxes back to equilibrium shortly after
the perturbation has been applied or the perturbation oscillates
showing quasi-periodic or even chaotic behavior. In the second case the
full equilibrium
is never reached. The system always keeps information about the initial
perturbation and releases it from time to time reproducing the shape of
initial perturbation. This behavior is typical for  unitary system in
finite
volume. General arguments show that time evolution of  unitary system
should show
the so-called Poincar\'e recurrences although the recurrence time may
be
extremely large, this time is estimated as $e^S$ where $S$ is entropy
of the system.
The information is thus never lost in finite volume -- the clear
manifestation
of the unitary evolution. The exponential decay  may happen only in
infinite volume or if the system is in finite volume but is
non-unitary.
The obvious example of the latter is black hole whose semiclassical
behavior
appears to violate unitarity.

\subsection{Relaxation in black hole: quasi-normal modes}
We consider (2+1)-dimensional BTZ black hole with metric given by
\be
ds^2=-\sinh^2y~ dt^2+dy^2+\cosh^2y~ d\phi^2~~,
\lb{1}
\ee
where for simplicity we consider non-rotating black hole and set the
size of
the horizon $r_+=L$ and AdS radius $l=1$. The coordinate $\phi$ is
periodic with period $L$ so
that the boundary has topology of cylinder and $L$ sets the finite size
for
the boundary system. $L$ is related to the mass of BTZ black hole as
$L=\sqrt{MG}$.
A bulk perturbation $\Phi_{(m,s)}$ of mass $m$ and spin $s$
should satisfy the quasi-normal boundary condition, i.e. it should be
in-going
at the horizon and have vanishing flux at the infinity. The latter
condition
comes from the fact that in the asymptotically AdS space-times the
effective
radial potential is growing at infinity so that there can  be no
propagating modes as well as no leakage of the energy through the
boundary.
The relevant radial equation takes the form of the hypergeometric
equation
which is exactly solvable. The  quasi-normal modes in general fall into
two
sets \cite{C-B,BSS2}
\begin{eqnarray}\label{2}
&&\omega={2\pi\over L}{\tt l}-4\pi i T_L(n+\bar h)\nonumber\\
&&\omega=-{2\pi\over L}{\tt l}-4\pi i T_R(n+h)~,~~ {\tt l}\in{\bf Z} ~
,~~ n\in{\bf N}
\end{eqnarray}
where the left- and right-temperatures $T_L=T_R=1/2\pi$ and $(h,
\bar{h})$
have the meaning of the conformal weights of the dual operator
${\mathcal
  O}_{(h,\bar{h})}$ corresponding to the bulk perturbation
$\Phi_{(m,s)}$, with $h+\bar{h}=\Delta (m), \ h-\bar{h}=s$ and
$\Delta(m)$ is
determined in terms of the mass $m$.

The appearance of complex modes (\ref{2}) does not come as a surprise
--
the quasi-normal boundary conditions are  dissipative in nature, they
say that
the perturbation once created should leave the region through all
possible
boundaries. Since no leakage of energy happens through the spatial
infinity
all the dissipation goes through horizon.
For comparison, in the case of global anti-de Sitter space the horizon
and
respectively the quasi-normal modes are absent. But, instead, one can
define
the normalizable modes by imposing the Dirichlet condition at infinity
as well
as regularity in the origin. They
form a discrete set of real frequencies \cite{Balasubramanian:1998sn}
\be
\omega=2\pi {\tt l}/L+4\pi(n+h)/L~,~~{\tt l}\in {\bf Z}~, ~~n\in {\bf
N}~~
\lb{3}
\ee
where the size of the boundary is also set to be $L$ as in the black
hole case.

A simple way to understand why black hole is not characterized by a set
of real
frequencies like (\ref{3}) is to observe that the perturbation
effectively
propagates in the infinite volume in the case of black hole. Indeed,
near
horizon the wave propagates freely in the so-called optical metric
defined as
\be
ds^2=\sinh^2y ~ds^2_{\tt opt}~~.
\lb{3'}
\ee
The distance to horizon in this metric is measured in terms of
coordinate
$z=\int dy/\sinh(y)$. Obviously it diverges as $y$ approaches horizon
at $y=0$.
Thus, the perturbation sees  effectively infinite volume and thus can
not be
characterized by a discrete set of real frequencies. We stress that it
is the
presence of horizon which creates this effective infinite volume and
eventually leads to appearance of the complex frequencies (\ref{2}).

\subsection{Relaxation in two-dimensional Conformal Field Theory}
The thermal state of the black hole in the bulk corresponds to  the
thermal
state  on the CFT side. In fact, the boundary CFT factorizes on left-
and
right-moving sectors with temperature $T_L$ and $T_R$ respectively. The
bulk
perturbation corresponds to perturbing the thermal field theory state
with
operator ${\mathcal O}_{(h,\bar{h})}$. The further evolution of the
system is
described by the so-called Linear Response Theory (see \cite{Fetter}).
According to this theory
one has to look at the time evolution of the perturbation itself.
More precisely, the relevant information is contained in the retarded
correlation function of the perturbation at the moments $t$ and $t=0$
(when the perturbation has been first applied). Since the perturbation
is
considered to be small the main evolution is still governed by the
unperturbed
Hamiltonian acting on the thermal state so that the correlation
function is the
thermal function at temperature $T$.
Thus, the analysis boils down
to the study of the thermal 2-point function of certain
conformal operators. Such a function should be doubly periodic: with
period
$1/T$ in the direction of the Euclidean time and with period $L$ in the
direction of the compact coordinate $\phi$. This can be first
calculated as a
2-point function on the Euclidean torus and then analytically continued
to the
real time.

\subsubsection{Large $L$/Small $T$ universality}
In general the correlation function on torus can be  rather complicated
since
its form is not fixed by the conformal symmetry. The conformal symmetry
however
may help to deduce the universal form of the 2-point function in two
special
cases: when size $L$ of the system is infinite (temperature $T$ is kept
finite)  and when inverse temperature is infinite (the size $L$ is
finite).
The universal form of the (real time) 2-point function  in the first
case is
\be
\< {\cal{O}}(t,\phi ){\cal{O}}(0,0 )\>
={(\pi T)^{2(h+\bar{h})} \over (\sinh \pi T (\phi-t
))^{2h}(\sinh \pi T (\phi+t ))^{2\bar{h}}}~~ \lb{31} \ee
which for large $t$ decays exponentially as
$e^{-2\pi T(h+\bar{h}) t}$. The information about the perturbation is
thus
lost after characteristic time set by the inverse temperature. It is
clear
that this happens because in  infinite volume the information may
dissipate
to infinity. In the second case correlator
\be
\<{\cal{O}}(t,\phi ){\cal{O}} (0,0 )\>
={(\pi/iL)^{2(h+\bar{h})} \over
(\sin \frac{\pi}{L}  (t+\phi ))^{2h}(\sin \frac{\pi}{L}  (t-\phi
))^{2\bar{h}}}~~
\lb{41}
\ee
has the oscillatory behavior. Notice that the oscillatory behavior in
the second case should have been expected since the system lives on
circle. Perturbation once created at the moment $t=0$ at  point
$\phi=0$ travels around the circle with the speed of light and comes
back to
the same point at $t=L$. Thus, the information about the perturbation
is never
lost. The correlation function (\ref{41}) as  function of
time  represents a series of singular peaks concentrated at $t=\pm
\phi+nL$, $n\in
{\bf N}$. In fact, this behavior should be typical for any system with
unitary
evolution in  finite volume.

It is interesting to see what happens in the intermediate regime when
both $L$
and $1/T$ are kept finite. In this case the behavior of the correlation
functions is not universal, may depend on the (self)interaction in the
system
and is known only in some cases.

\subsubsection{Intermediate regime: (quasi-)periodicity and unitarity}
As an example of a system in the intermediate regime when both $L$ and
$T$ are
kept fixed we consider the free fermions for which the correlation
function
on the torus is known explicitly (e.g.
\cite{DiFrancesco:nk}). The real   time correlation function is
\be
\<\psi (w)\psi(0)\>_{\nu}={\theta_\nu (wT |i
LT)\partial_{z}\theta_{1}(0|LT)\over
\theta_{\nu}(0|iLT)\theta_1 (wT |i LT)}~~,
\lb{5}
\ee
were $w=i(t+\phi)$ and $\nu$ characterizes the
boundary conditions for $\psi (w)$. For finite temperature boundary
conditions we have $\nu=3,4$. Using the properties of
$\theta$-functions, it is then easy to see that (\ref{5})
is invariant under shifts $w\rightarrow w+1/T$
and $w\rightarrow w+iL$.
It is then obvious that the resulting real time correlator
(\ref{5}) is a {\it periodic} function of $t$ with period $L$. Zeros of
the
theta function $\theta_1(wT|iLT)$ are known \cite{DiFrancesco:nk}
to lie at $w=m/T +inL$, where $m,n$ are
arbitrary relative integers. Therefore, for real time $t$, the
correlation
function (\ref{5}) is a sequence of singular peaks located at
$(t+\phi)=nL$. Using the standard representation \cite{DiFrancesco:nk}
of the $\theta$-functions, we also
find that in the limit $LT\rightarrow \infty$ the correlation function
(\ref{5}) approaches
the left-moving part of (\ref{3}) with $h=1/2$
that exponentially decays with time,
\be
\<\psi (w)\psi(0)\>_{3(4)}={\pi T\over 4\sinh{\pi T(t+\phi)}}[1\pm
2e^{-\pi LT}\cosh 2\pi T(t+\phi)+..]~~.
\lb{61}
\ee
In the opposite limit, when $LT\rightarrow 0$,
it approaches the oscillating function (\ref{41}).
A natural question is how the asymptotic behavior (\ref{61}) when size
of the system is taken
to infinity can be consistent with the periodicity, $t\rightarrow t+L$,
of the correlation function (\ref{5}) at any finite $L$?
In order to answer this question we have to observe that there are two
different time scales in the game. The first time scale is set by the
inverse
temperature $\tau_1=1/T$ and  while the second time scale is associated
with the size of the system $\tau_2=L$. When $L$ is taken to infinity
we
have that
$\tau_2\gg\tau_1$. Now, when the time $t$ is of the order of $\tau_1$
but much
less than $\tau_2$ the asymptotic expansion (\ref{61}) takes place. The
corrections to the leading term are multiplied by the factor $e^{-\pi
LT}$ and
are small. The 2-point function thus is exponentially decaying in this
regime.
It seems that the system has almost lost   information about the
initial
perturbation (at $t=0$). But it is not the case: as time goes on and
approaches the second time scale $t\sim \tau_2$ the corrections to the
leading term in (\ref{61}) become important and the system starts to
collect
its memory about the initial perturbation. The information is
completely
recovered as $t=\tau_2$ and the time-periodicity restores.
This example is instructive. Provided the two scales $\tau_1$ and
$\tau_2$ are
widely separated, $\tau_1\ll\tau_2$, the system would seem to relax
exponentially fast back to thermal equilibrium for the moments of time
$t$ such that $t\gg\tau_1$ but $t\ll\tau_2$. Observing the system
during these $t$ we would  have concluded that the information about
the initial
perturbation had been lost completely  and that the system  was showing
a
non-unitary behavior. The unitarity however completely restores if we
wait
long enough until $t$ gets close to the second time scale $\tau_2$.
In general we should expect that unitary system in finite volume is
characterized by a set of periods so that its time evolution is
quasi-periodic.

\subsection{CFT$_2$ dual to AdS$_3$}

\subsubsection{Correlation functions}
As an example of a strongly coupled
theory we consider the supersymmetric conformal field theory dual to
string
theory on AdS$_3$. This theory describes the low energy excitations of
a large number of D1- and D5-branes \cite{Aharony:1999ti}. It can be
interpreted as a gas of
strings that wind around a circle of length $L$ with target space
$T^4$.
The individual strings can be simply- or multiply wound such that the
total
winding number is ${\tt k}=\frac{c}{6}$, where $c\gg 1$ is the central
charge.
The parameter $\tt k$ plays the role of N in the usual terminology of
large N CFT.

According to the  prescription (see \cite{Aharony:1999ti}), each AdS
space
which asymptotically approaches the given
two-dimensional manifold should contribute
to the calculation, and one thus has to sum over all such spaces.
In the case of interest, the two-manifold is a torus $(\tau_E, \phi )$,
and $\beta=1/T$ and $L$ are the respective periods.
There exist two obvious AdS spaces which approach the torus
asymptotically.
The first is the BTZ black hole  and
the second is the so-called thermal AdS space,
corresponding to anti-de Sitter space filled with thermal radiation.
Both spaces can be represented (see \cite{Carlip:1994gc})
as a quotient of three dimensional
hyperbolic space $H^3$, with line element
\be
ds^2={l^2\over y^2}(dzd\bar{z}+dy^2)~~~y>0~~.
\lb{7}
\ee
In both cases, the boundary of
the three-dimensional space is a rectangular torus with
periods $L$ and $1/T$. Two configurations
(thermal AdS and the BTZ black hole) are T-dual to
each other, and are obtained by the interchange of the coordinates
$\tau_E \leftrightarrow \phi$ and $L \leftrightarrow 1/T$ on the torus.
In fact there is a whole $SL(2,{\bf Z})$ family of spaces which are
quotients
of the hyperbolic space.

In order to find correlation
function of the dual conformal operators, one has to solve the
respective bulk field equations subject to Dirichlet boundary
condition, substitute the solution into the action and
differentiate the action twice with respect to the boundary value
of the field. The boundary field thus plays the role of the source
for the dual operator ${\cal{O}}_{(h,\bar{h})}$.
This way one can obtain the boundary CFT correlation function for each
member
of the family of asymptotically AdS spaces.
The total correlation function is then given by the sum over all
$SL(2,{\bf
  Z})$ family with appropriate weight. Let us however first
consider the contribution of only two
terms  \cite{Maldacena:2001kr}
\be
\<{\mathcal{O}}(t,\phi) {\mathcal{O}} (0,0)\> \lb{8}
\simeq
e^{-S_{\tt BTZ}}\<{\mathcal{O}} ~{\mathcal{O}}'\>_{\tt BTZ}+
e^{-S_{\tt AdS}}\<{\mathcal{O}} ~{\mathcal{O}}'\>_{\tt AdS}~~,
\ee
where
$S_{\tt BTZ}=-{{\tt k} \pi LT/2}$ and $S_{\tt AdS}=-{{\tt k}  \pi
/2LT}$
are Euclidean action of the
BTZ black hole and thermal AdS$_3$, respectively
\cite{Maldacena:1998bw}.
On the Euclidean torus $\< \ \ \>_{\tt BTZ}$ and $\< \ \ \>_{\tt AdS}$
are T-dual to each other.
Their exact form can be computed explicitly \cite{Corrfunct}.
After the analytical continuation $\tau_E=it$, the BTZ contribution
\be
&&\<{\cal{O}} (t,\phi ){\cal{O}} (0,0)\>_{\tt BTZ} =\lb{10}
\sum_n{1\over
(\sinh \frac{\pi}{\beta} (\phi-t+nL ))^{\Delta}
(\sinh \frac{\pi}{\beta} (t+\phi+nL ))^{\Delta}}~~
\ee
exponentially decays with time.
The result for the thermal AdS
\be
&&\<{\cal{O}} (t,\phi ){\cal{O}} (0,0)\>_{\tt AdS}
=\sum_n{1\over
(\sin {\pi\over L}  (t+\phi +i\beta n))^{\Delta}
(\sin {\pi\over L}  (t-\phi +i\beta n))^{\Delta}} ~~
\lb{11}
\ee
is periodic in time with period $L$. It represents  a periodic
sequence of singular peaks at $t\pm\phi =nL$.

Thus, the total 2-point function (\ref{8}) has two contributions: one
is exponentially
decaying and another is oscillating.  So that (\ref{8}) is {\it not} a
quasi-periodic function  of time $t$.
This can be re-formulated   in terms of the poles in the momentum
representation of 2-point function (see \cite{BSS2} and
\cite{Danielsson:1999zt}). The poles of $\< \ \ \>_{\tt BTZ}$
are exactly the complex quasi-normal modes (\ref{2}) while that of
 $\< \ \ \>_{\tt AdS}$  are the real normalizable modes (\ref{3}).
Depending on the value of
$LT$, one of the two terms in (\ref{8}) dominates
\cite{Maldacena:1998bw}.
For high temperature ($LT$ is large)
the BTZ is dominating, while at low temperature  ($LT$ is small)
the thermal AdS is dominant. The transition between the two regimes
occurs at $1/T=L$. In terms of the gravitational physics, this
corresponds to the
Hawking-Page phase transition \cite{Hawking:1982dh}.
This is a sharp transition for large $\tt k$, which is the case when
the supergravity description is valid.
The Hawking-Page transition is thus a transition between oscillatory
relaxation at low temperature and exponential decay at high temperature
\cite{BSS}.

Whether including sum over $SL(2,{\bf Z})$ in (\ref{8}) we can get a
quasi-periodic
result is discussed in the next subsection.

\subsubsection{Can  sum over $SL(2,{\bf Z})$ family produce
quasi-periodic result?}
In the AdS/CFT correspondence we are instructed to sum over all
possible AdS
metrics which approach same boundary manifold at infinity. In the case
at hand
the boundary manifold is two-dimensional torus $(\phi,\tau_E)$ with
periodicities $(\phi,\tau_E)\rightarrow (\phi+Ln,\tau_E+\beta m)$,
$n,m\in Z$.
The complex holomorphic coordinate on the torus is $w=\phi+i\tau_E$.
The torus
is characterized by the modular parameter $\tau=i\beta/L$. The
$SL(2,{\bf Z})$
modular transformations act as
\be
\tau\rightarrow \tau'={a\tau+b\over c\tau +d}~,~~w\rightarrow
w'={w\over
  c\tau+d}~,~~ad-bc=1~~,
\lb{s1}
\ee
where group parameters $a,b,c,d$ are integers. In fact we should be
interested
in the $SL(2,{\bf Z})/{\bf Z}$ transformations, i.e. modulo the
parabolic group transformations
$(a,b)\rightarrow (a+c,b+d)$. These transformations are completely
determined
by pairs of relatively prime $(c,d)$. In the $SL(2,{\bf Z})$ family the choice
$(a=1,b=c=0,d=1)$
corresponds to the thermal $AdS_3$ while the choice
$(a=0,b=1,c=-1,d=0)$
describes BTZ  black hole. The gravitational action for a AdS metric
asymptotically approaching the torus characterized by modular parameter
$\tau'$ takes the form \cite{Maldacena:1998bw}
\be
S({\tau'})={\pi {\tt k} i\over 2}\left[{a\tau+b\over
c\tau+d}-{a\bar{\tau}+b\over
    c\bar{\tau}+d}\right]=-\pi {\tt k}{\beta L\over
c^2\beta^2+d^2L^2}~~.
\lb{s2}
\ee
Applying now AdS/CFT rules  for computing the  2-point function on the
torus
we have to sum over the $SL(2,{\bf Z})/{\bf Z}$ family with appropriate
weight \cite{Porrati},
\be
\<{\cal O}(w,\bar{w}){\cal O}(0,0)\>_{\tt SL(2,{\bf Z})}=
\sum_{(c,d)}\sum_{n\in
Z}{e^{-S({\tau'})}\over |c\tau +d|^{2\Delta}}{(\pi/L)^{2\Delta} \over
|\sin{\pi ({w'\over
    L}+n\tau')}|^{2\Delta}}~~,
\lb{s3}
\ee
where $w'$ and $\tau'$ are defined in (\ref{s1}) and for simplicity we
take
$h=\bar{h}=\Delta/2$. An important question is what to choose for the
weights
in (\ref{s3}). Explicit calculation \cite{Farey} for the field theory
elliptic genus
shows that the weight of each geometry might be rather complicated and
is not
just the Euclidean action. In (\ref{s3}) however we made this simplest
choice
in order to simplify our analysis.

Since we are  interested in the real time correlators, we should
analytically continue  $\tau_E=it$. This gives the substitution
$w=\phi-t$,
$\bar{w}=\phi+t$ in (\ref{s3}). The correlation function (\ref{s3})
then can be
written in the form  the
\be
&&\<{\cal O}(t,\phi){\cal O}(0,0)\>_{\tt SL(2,{\bf Z})}=\nonumber \\
&&\sum_{(c,d)}\sum_{n\in
Z}e^{{\pi {\tt k}\beta L\over c^2\beta^2+d^2L^2}}\left[{L^2\over
c^2\beta^2+d^2L^2}\right]^\Delta {(\pi/L)^{2\Delta}\over (\sinh
\pi(x^+_{n(c,d)})\sinh\pi(x^-_{n(c,d)}))^\Delta}~~,
\lb{s4}
\ee
where
\be
&&x^\pm_{n(c,d)}={\beta\over c^2\beta^2+d^2L^2}(c(t\pm\phi)\pm
nL)\nonumber \\
&&-i{L\over
  c^2\beta^2+d^2L^2}(d(\phi\pm t)+n{(ac\beta^2+bdL^2)\over L})~~.
\lb{s5}
\ee
For large $c$ or $d$  the quantities $x^\pm_{n(c,d)}$ accumulate near zero so that
convergence of (\ref{s4}) is not obvious. (A particular
divergent contribution is due to terms with $c=d-1$, $a=b=1$.)
Some regularization  of (\ref{s4}) may be needed. A possible one is
to replace $x^\pm_{n(c,d)}\rightarrow x^\pm_{n(c,d)}\pm i\gamma$ where $\gamma$ is real.
Below some regularization of this sort is assumed.
In the $SL(2,{\bf Z})$ family the configuration with $c=0$ is thermal
AdS for which
correlation function (\ref{s4}) is oscillating (see (\ref{11}))  while
the
ones with $c\neq 0$ are black holes  for which the correlation function
exponentially decays as in (\ref{10}).

What can we say about the temporal behavior of  sum  (\ref{s4})? Can it
be
quasi-periodic even though  it is built out of the exponentially
decaying
pieces? Answering these questions it is instructive to look at the
large $t$ behavior
of (\ref{s4}) at fixed $n$. We find (see also \cite{Porrati} for a
related analysis)
\be
\<{\cal O}(t,\phi){\cal O}(0,0)\>_{\tt SL(2,{\bf Z})}=\sum_{(c,d)}
e^{{\pi
    {\tt k}\beta L\over c^2\beta^2+d^2L^2}}\left[{\pi^2\over
c^2\beta^2+d^2L^2}\right]^\Delta e^{-{2\Delta \beta \pi |c|t\over
c^2\beta^2+d^2L^2}}~~.
\lb{s6}
\ee
Each term in (\ref{s6}) is exponentially decaying. However, the
characteristic
decay time  for  term characterized by pair $(c,d)$ becomes arbitrary
large
for large $c$ and $d$. This means that such terms become relevant at
later and later times. In fact it is an indication of that the sum
(\ref{s6})
actually decays with time slower than an exponent.
On the other hand, as was observed in \cite{Porrati},  at certain
critical
time $t_c={{\tt  k}L\over 2\Delta |c|}$ the decaying exponential factor
compensates the exponent of
action in (\ref{s6}) so that for time $t\geq {{\tt k}L\over 2\Delta}$
there is no
suppression in  (\ref{s6}) and all terms in
the sum  are equally  important. In \cite{Porrati} this was interpreted
as
a signal of breakdown of the semiclassical approximation.

It is instructive  to analyze those issues in a simple example.
Consider the sum
\be
I(t)=\sum_{n=1}^\infty {1\over n^2}\ e^{a^2/n^2}\ e^{-t/n}
\lb{s7}
\ee
which consists of exponentially decaying terms but the characteristic
decay time
grows with $n$. As a result, the sum decays slower than  exponential
function.
In fact, as we show below, it falls off by power law.
At $t=a^2/n$ the two exponential functions in (\ref{s7}) compensate
each other
so that for $t>a^2$ all terms in the sum are equally contributing.
One might worry whether sum (\ref{s7}) is actually convergent at
$t=a^2$. Obviously, this  is a
false alarm -- sum (\ref{s7}) is
convergent for any $t$ due to factor $1/n^2$.
In order to make all these points more transparent  we  approximate the
sum (\ref{s7}) by integral
\be
I(t)\simeq \int_1^\infty {dn\over n^2}e^{a^2/n^2}e^{-t/n}=
{\sqrt{\pi}\over 2}i[\Phi(-i{t\over 2a})-\Phi(i(a-{t\over 2a}))]= {a^2
\over
  t}+O({a^2e^{a^2}\over t}e^{-t})~,
\lb{s8}
\ee
where in the last passage we assumed that $t\gg 2a^2$ and
approximation for
the error function $\Phi(z)$ of large (complex)
argument was used.
Obviously, there is nothing special happening at $t=a^2$. For large
$t\gg
2a^2$ the sum
$I(t)$ falls off by power law as anticipated.

Returning to our sum (\ref{s6}) we may expect that it actually falls
off by a
power law similarly to the sum $I(t)$. In order to see this we can use
similar
trick and replace infinite sum $\sum_{(c,d)}$ by integral $\int dc \
dd$. Obviously, this will overestimate the actual sum and hence give an upper bound on it.
Considering
$c$ and $d$ as continuous variables it is useful to make a
transformation to
``polar'' coordinate variables $(\rho,\varphi)$
\be
c={\rho\over \beta}\cos\varphi~,~~d={\rho\over L}\sin\varphi~~.
\lb{s9}
\ee
Then we get that
\be
&&\<{\cal O}(t,\phi){\cal O}(0,0)\>_{\tt SL(2,{\bf
Z})}\simeq{2\pi^{2\Delta}\over
 L \beta}\int_R^\infty {d\rho\over \rho^{2\Delta-1}}e^{\pi {\tt k}\beta
L\over
    \rho^2}\int_0^{\pi/2}d\varphi \ e^{-{2\Delta \pi t\over
\rho}\cos\varphi}~~,
  \nonumber \\
&&\int_0^{\pi/2}d\varphi \ e^{-{2\Delta \pi t\over
\rho}\cos\varphi}={\pi\over
  2}(I_0({2\Delta \pi t\over \rho})-{\bf L}_0({2\Delta \pi t\over
  \rho}))~~,
\lb{s10}
\ee
where $R\simeq \sqrt{\beta^2+L^2}$. The $\rho$ integral is peaked at
lower
limit and we can approximate $e^{\pi {\tt k}\beta L\over
    \rho^2}\sim e^{\pi {\tt k} \beta L\over R^2}$. The integration over
$\rho$  in (\ref{s10}) then
  can be performed explicitly for any integer $\Delta$. The result is
  expressed in terms of Bessel $I_k(z)$ and Struve ${\bf L}_k(z)$
functions. For $\Delta=2$ in
  particular we have
\be
\<{\cal O}(t,\phi){\cal O}(0,0)\>_{\tt SL(2,{\bf Z})}\simeq {1\over
  L\beta}{e^{\pi{\tt k}\beta L \over R^2}\over 4\pi
  t R}[I_1({4\pi t\over R})-{\bf L}_1({4\pi t\over R})]\sim
{1\over t}{1 \over \beta L R}e^{\pi {\tt k} \beta L/(\beta^2+L^2)}~~,
\lb{s11}
\ee
where in the last passage we take the limit of large $t$. In fact, the
analysis shows
that  the power
law $1/t$ is universal large $t$ behavior of (\ref{s10}) for all values
$\Delta \geq 2$.

We see that   sum over pairs $(c,d)$ produces
correlation function which decays by a power law.
It is an improvement over the exponential decay of each
individual contribution (due to black holes) in the sum. This, however,
does
not produce a quasi-periodic correlation function.

\subsubsection{Unitarity: boundary theory and black hole}
Thus, the AdS/CFT correspondence predicts that the relaxation in
CFT dual to gravity on AdS$_3$ is
 combination of oscillating and  decaying functions.
Thus, the information about initial perturbation seems to be inevitably
lost
in the boundary system. Since the latter lives in finite volume
(circle of size $L$) and is supposed to be unitary there must be a way
to
resolve this apparent contradiction.
The resolution was suggested in \cite{BSS} where it was noted that
at finite $k$ there should exist another scale in the system which is
additional to and much large than $L$. This scale appears due to
the fact that in the dual CFT at high temperature
the typical configuration consists of multiply wound strings  which
effectively propagate in a much bigger volume, $L_{\tt eff}\sim {\tt
k}L$.
The gravity/CFT duality however is valid in the limit of infinite $\tt
k$
in which this second  scale becomes infinite. So that the exponential
relaxation corresponds to infinite effective size $L_{\tt eff}$
that is in complete agreement with the general arguments. At finite
$\tt k$
the scale $L_{\tt eff}$ would be  finite and the correlation function
is expected to be quasi-periodic with two periods $1/L$ and $1/L_{\tt
eff}$.
The transition of this quasi-periodic function to combination of
exponentially
decaying and oscillating functions when $L_{\tt eff}$ is infinite then
should
be  similar to what we have observed in the case of free fermions when
$L$
was taken to infinity.

On the gravity side the question  of which type of relaxation occurs in
the
system is related to one of the most fundamental problems in physics --
the problem of
information loss and black hole unitarity.
The unitarity problem was suggested to be resolved within the AdS/CFT
correspondence \cite{Maldacena:2001kr}. Indeed, since the theory on the
boundary is unitary  it should
be possible to reformulate all processes happening in the bulk of black
hole space-time
on the intrinsically  unitary language of the boundary CFT.
In particular it was suggested \cite{Maldacena:2001kr} that in order to
restore unitarity of physics in the bulk and reproduce the expected
unitary behavior
of boundary theory  we have to take into account the topological
diversity of gravitational instantons that asymptotic to given boundary
manifold. If worked this way the black hole unitarity would be resolved
within
semiclassical gravity appropriately redefined to account for all
possible topologies.
Alternatively and perhaps more traditionally the unitarity might be
expected
to restore as a result of fundamentally non-perturbative effects of
Quantum
Gravity. On the present stage of our understanding of Quantum Gravity
this
second way would inevitably involve making some guesses about
the non-perturbative behavior of black hole.

The relaxation phenomenon gives us adequate language
for analysis of the problem of black hole unitarity.
That relaxation of black hole is characterized by a set of complex
frequencies
(quasi-normal modes) is mathematically precise formulation of the lack
of
unitarity in the semiclassical description of black holes.
The loss of information in semiclassical black hole is indeed visible
on the CFT side.
It is encoded in the exponentially decaying contribution to the 2-point
correlation function. For the CFT itself this however is not a problem.
As we discussed above the finite size unitarity  restores at finite
value
of $\tt k$. This however goes beyond the limits where the gravity/CFT
duality
is formulated. Assuming that the duality can be extended to finite
$\tt k$ an important question arises: {\it What would be the gravity
counter-part
  of the duality at finite $\tt k$?} Obviously, it can not be a
semiclassical
black hole or ensemble of topologically distinct  semiclassical black
holes.
The black hole horizon should be somehow removed so that the
complex quasi-normal modes (at infinite $\tt k$) would be replaced with
real (normal) modes when $\tt k$ is finite. This puts us on the second
track
of resolving the unitarity problem: within a non-perturbative treatment
of
Quantum Gravity.
So that we should start with making our best guess about the
nonsemiclassical
description of black hole horizons.

\section{Wormhole modification of near-horizon geometry}
\setcounter{equation}0
Before passing to our proposal for the nonsemiclassical black hole let
us pause
for a moment and discuss another proposal made almost twenty years ago
by 't
Hooft and called the ``brick wall'' \cite{tHooft} (see also
\cite{R.Mann}).
It was an attempt to explain the entropy of black hole as   entropy of
thermal atmosphere of
particles outside the black hole horizon. In D space-time dimensions
the
free energy  of thermal gas at temperature $T$ in finite volume
$V_{D-1}$
takes the form
\be
F=-\pi^{-D/2} \Gamma({D\over 2})\ \zeta (D)T^DV_{D-1}~~.
\lb{13}
\ee
Near horizon  the appropriate volume $V_{D-1}$ is defined
in the optical metric (\ref{3'}) and is infinite. This means that the
spectrum
of field excitations is continuous and the entropy is infinite.
In order to regularize it
't Hooft suggested to cut the region just outside horizon by
introducing  boundary
at small distance $\epsilon$ from horizon  and imposing there Dirichlet
boundary condition. This procedure has two important consequences:
the spectrum now becomes discrete, $\omega_n\sim \pi n/L_{\tt opt}$,
$n\in Z$
and $L_{\tt opt}=\ln{1\over \epsilon}$ is the ``size'' of the finite
region;
and the entropy which can be computed from (\ref{13}) using standard
formula
$S=-{\partial F\over \partial T}$ becomes finite. Moreover, due to
remarkable
property of the optical volume $V_{D-1}\sim {A\over \epsilon^{D-2}}$
the entropy calculated this way  is proportional to the horizon area
$A$,
$S\sim {A\over \epsilon^{D-2}}$. The entropy diverges when
$\epsilon$ is taken to zero and there was a lot of discussion in the
literature in the 90's what this divergence should mean
\cite{Brickwall}.
For our story it is important to note that once horizon has been
removed
the system now lives in finite optical volume and, most importantly,
the complex quasi-normal modes disappear. This is exactly what we need
in order
to restore  unitarity in black hole \cite {Barbon}.
The brick wall model however is an artificial way to
regularize the otherwise smooth black hole geometry. It can be
considered
as rather crude way of presenting  the unknown non-perturbative
Planckian
physics.

By our earlier (unpublished) work  \cite{SS1} there however exists
a smooth way of changing the near horizon geometry
which would now look like a wormhole connecting two asymptotic regions
semiclassically separated by horizon. This modification of black hole
geometry
does the same job as the brick wall, i.e. leads to discrete spectrum
and finite
entropy, but does it in a smooth way. In the context of the black hole
relaxation
which is subject of the present study the wormhole modification has
some
attractive features absent in the brick wall model and in fact
describes the
expected unitary relaxation quite naturally. We study these issues in
the next sections.
Here we first introduce the modified geometry for the BTZ black hole
\cite{SS},
\be
ds^2=-(\sinh^2y+\lambda^2({\tt k}))~ dt^2+dy^2+\cosh^2y~ d\phi^2 ~~.
\lb{9}
\ee
The deformation parameter $\lambda({\tt k})$ is supposed to be some
function of the large N
parameter ${\tt k}$ such that it vanishes when ${\tt k}$ is infinite.
Concrete form of this function is discussed later in this Section.
The horizon located  at $y=0$ in classical BTZ black hole
disappears in metric (\ref{9}) if $\lambda$ is non-vanishing.
The whole geometry now is that of wormhole  with the second asymptotic
region at $y<0$.
The two asymptotic regions ($y>0$ and $y<0$) which were separated by
horizon in classical BTZ metric
(\ref{1}) are now connected through narrow throat and thus can
talk to each other  exchanging information.
The metric (\ref{9}) is still asymptotically AdS although it is no more
a constant
curvature space-time. The Ricci scalar
\be
R=-{2\over (
\sinh^2y+\lambda^2)^2}[\lambda^2+\lambda^4+3\sinh^4y+5\lambda^2 \sinh^2y ]
\lb{12}
\ee
approaches value $-6$ at infinite $y$ and $-2(1/\lambda^2+1)$ at $y=0$
where the
horizon used to stay. Notice that the parameter $\lambda$ should
account for
the quantum Planckian effects. The curvature at the throat which
replaced
horizon is thus of the Planckian order.
The metric (\ref{9}) is  an
example that shows that horizon as causally special set in  space-time
can be seen as a place which  accumulates the
quantum effects so that under the small deformation there appears
Planckian
scale curvature concentrated in the  Planck size region.

The metric (\ref{9}) can be brought to the usual Schwarzschild like
form
introducing the radial coordinate $r=\cosh y$,
\be
ds^2=-(r^2-1+\lambda^2)dt^2+{dr^2\over r^2-1}+r^2d\phi^2~~.
\lb{r}
\ee
In the semiclassical case ($\lambda=0$) it was possible to extend
the metric to include the region where $r^2<1$ that was joint to the
region
$r^2>1$ along the light-like horizon $r=1$. In the non-semiclassical
case
(\ref{r}) there appears an intermediate region $(1-\lambda^2<r^2<1)$
where
the signature becomes $(- \ - \ +)$  (i.e. it is  spacetime with two
time-like
coordinates) and which can not be extended
neither to region $r^2\geq 1$ nor to region $r^2\leq 1-\lambda^2$. The
latter two
regions are thus disconnected from each other and present two different
space-times.

Let us now illustrate our point that the entropy of the thermal gas in
the
metric (\ref{9}) has finite entropy. Indeed, the  optical volume
\be
V_{\tt opt}=2\int_0^\infty{dy \cosh y\over \sinh^2y +\lambda^2({\tt
k})} ~
L=\pi \lambda^{-1}({\tt k}) A~~,
\lb{14}
\ee
where $A=L$ is the ``area'' of horizon, is now finite. Applying now
formula
(\ref{13}) for $D=3$ and taking into account that the Hawking
temperature (we
take the classical value for the temperature) is $T=2\pi$ we find that
the
entropy
\be
S=6\pi^2 \zeta(3)\lambda^{-1} ({\tt k}) A
\lb{15}
\ee
is finite. If there are $N$ species of particles all of them should be
taken
into account so that the entropy (\ref{15}) would be proportional to
$N$.
In principle, playing with two free parameters $N$ and $\lambda$ we can
easily
match the entropy (\ref{15}) with the Bekenstein-Hawking entropy of BTZ
black
hole. This  however is not our primary goal in this paper.

It is clear that the spectrum of field excitations in the metric
(\ref{9}) should be
discrete. This is just because the size of the space-time in the
optical
metric measured from one boundary to another is finite
\be
L_{\tt opt}=2\int_0^\infty {dy\over \sqrt{\sinh^2y+\lambda^2}}=
2{\bf    K}(\sqrt{1-\lambda^2})=2\ln{4\over
\lambda}+O(\lambda^2\ln\lambda)~~,
\lb{16}
\ee
where ${\bf K}(z)$ is elliptic integral and we used one of its
expansions.
The expected spectrum in the limit of small $\lambda$ then reads
\be
\omega_n\simeq{\pi\ n\over L_{\tt opt}}\simeq{\pi \ n\over 2\ln{4\over
    \lambda}}~,~~n \in {\bf Z}~~
\lb{17}
\ee
for large $n$, that agrees with the earlier mode calculation in
\cite{Siopsis:2004gj}.
These are the normal frequencies in the metric (\ref{9}).
Comparing the two approaches, the brick wall and the wormhole
modification,
we see that there exists a correspondence between them provided we make
a
substitution $\lambda\leftrightarrow \epsilon$.

We finish this section with discussion on the possible form of the
deformation parameter $\lambda$ as function of the large N parameter
$\tt k$.
One obvious choice is $\lambda=a/{\tt k}$ where $a$ is some unknown
factor.
The advantage of this choice is that the entropy (\ref{15}) takes the
form
\be
S=\# \ {\tt k} A~~.
\lb{18}
\ee
The numerical factor in front of (\ref{18}) can be chosen  (by changing
parameter $a$) in a way that
(\ref{18}) exactly reproduces the classical Bekenstein-Hawking entropy.
The normal frequencies (\ref{17}) then would scale as $\sim 1/\ln{\tt
k}$.

Another choice is $\lambda({\tt k})\sim e^{-\tt k}$. Recalling relation
between
$\tt k$ and the Newton constant, $k=1/4G$, the wormhole modification
appears
as non-perturbative Quantum Gravity effect, $\lambda\sim e^{-1/4G}$.
The normal modes then scale as $1/{\tt k}$ and, choosing $\lambda\sim
e^{-{\tt
    k}L}$, we can identify
$L_{\tt opt}$ and $L_{\tt eff}$ introduced in section 2.3.3.
Notice that $S={\tt k}L$ is the entropy of BTZ black hole.
This choice seems to
be preferable since in this case  the geometry (\ref{9})  originates
in completely non-perturbative fashion.

\section{Applying the AdS/CFT rules: conformal anomaly}
The metric (\ref{9}) is asymptotically AdS and thus we can use the
AdS/CFT
rules and extract information about boundary theory from the asymptotic
behavior of the metric. In particular we can calculate the conformal
anomaly
in the boundary theory (see \cite{cft}, \cite{cmp} for more details).
For that we first introduce a new radial coordinate
$\rho=e^{-2y}$ and re-write the metric (\ref{9}) in the form
\be
&&ds^2={d\rho^2\over 4\rho^2}+{1\over \rho}g^{}_{ij}(x,\rho )dx^i
  dx^j~~, \nonumber \\
&&g_{ij}(x,\rho )=\sum_n g^{(n)}_{ij}(x)\rho^{2n}~~,
\lb{c1}
\ee
where $x^i=(t,\phi)$ are coordinates on the boundary. Clearly the
metric
(\ref{9})
takes the form (\ref{c1}) with
\be
&&g^{(0)}_{\phi\phi}=-g^{(0)}_{tt}=1~,~~\nonumber \\
&&g^{(2)}_{\phi\phi}={1\over
  2}~,~~g^{(2)}_{tt}={1\over 2}-\lambda^2({\tt k})~, \nonumber \\
&&g^{(4)}_{\phi\phi}=-g^{(4)}_{tt}={1\over 16}~~.
\lb{c2}
\ee
The vacuum expectation value of the boundary (quantum) stress tensor
can be calculated using the formula \cite{cmp}
\be
\<T_{ij}\>={2l\over 16\pi G}(g^{(2)}_{ij}-g^{(0)}_{ij} \Tr g^{(2)})~~.
\lb{c3}
\ee
In particular the trace of the stress tensor that represents the
conformal
anomaly in the boundary theory is given by the formula
\be
\Tr \<T\>=-{2l\over 16\pi G}\Tr g^{(2)}~~.
\lb{c4}
\ee
Substituting here the asymptotic expansion (\ref{c2}) for the metric
(\ref{9}) we
find
\be
\Tr \<T\>={c\over 24\pi}[-2\lambda^2]
\lb{c5}
\ee
for the conformal anomaly, where $c=3l/2G$ is the central charge,
$c=6{\tt k}$.
The interesting fact about this conformal anomaly is that it is
non-vanishing.
Indeed, the conformal anomaly in two dimensions is given by the
Ricci-scalar.
The two-dimensional space-time lying in the boundary of the
three-dimensional
space-time (\ref{9}) is flat. In the Euclidean case, when the Euclidean
time is
compactified, this space is two-dimensional torus. Since the
Ricci-scalar
is vanishing in this case we should not normally expect any conformal
anomaly
to appear.
Surprisingly, the above calculation gives us non-trivial anomaly
determined by
parameter $\lambda$ and this needs to be explained. The explanation we
can
offer is rather
simple. We suggest that the metric (\ref{9}) effectively describes
boundary theory
with mass gap determined by parameter $\lambda$\footnote{Alternative
interpretation of the trace anomaly
(\ref{c5}) is that it appears due to
non-vanishing two-dimensional cosmological constant proportional to $\lambda$.}.
Indeed, in this case
the
conformal anomaly can be built from a pair of dimension two quantities:
Ricci-scalar and the mass squared. For instance, a free massive scalar
field ($c=1$)
has conformal anomaly given by
\be
\Tr \<T\>={1\over 24\pi}[R-6m^2]~~.
\lb{c6}
\ee
Similar expression exists for free massive fermions. In flat
space-time, when
Ricci-scalar is vanishing, the conformal anomaly is given by the mass
squared only.
Comparison with (\ref{c5}) suggests that there is a mass gap $m\sim
\lambda$
in the boundary theory.
The exact proportionality coefficient can not be
determined from these arguments since the boundary theory is strongly
interacting
while the expression (\ref{c6}) is given for non-interacting scalar
field.
Recalling that $\lambda\sim e^{-{\tt k}}$ and ${\tt k}={1\over 4G}$ we
see that the appearance of the mass
gap is yet another non-perturbative effect  encoded in the shape of the
metric (\ref{9}).

\section{The quasi-periodicity: time scales and the spectrum}
Demonstrating the quasi-periodicity in the time evolution of   field
perturbations in the metric
(\ref{9}) we first notice that there exist two time  scales associated
with the metric (\ref{9}). The first one is given by the optical length
$L_{\tt
  opt}\sim \ln {1\over \lambda}$ and another is determined by the size
$\lambda$ of the throat. When $\lambda$ is small the two time scales
are widely
separated, $1/\lambda \gg \ln{1\over \lambda}$, so that we can talk
about
hierarchy of time scales.
The time scale $1/\lambda$ appears when we look at the metric
(\ref{9}) in the throat region ($y$ is close to zero) and find that it
is
basically flat with the ``throat time'' being rescaled with respect to
the
time $t$ at infinity as $t_{\tt thr}\sim\lambda \ t$. Assuming that
$\lambda\sim
e^{-{\tt k}}$ we find that time flows in the throat   extremely slow.
So that  processes which are rapidly changing with respect to the time
in the throat
are practically frozen as measured by clocks
at infinity.
It is exactly this property of the wormhole
geometry (\ref{9}) that makes  the black hole unitarity restored
in the long period of time set by $1/\lambda$ which is the maximal time
scale
in the system. This is the Poincar\'e recurrence time for the black
hole that
was missing in the semiclassical description.

Let us now turn on a field perturbation in the background of metric
(\ref{9})
and see what are the  frequencies which characterize the time evolution
of the
perturbation. For simplicity we consider minimally coupled  scalar
field with vanishing mass.
Making anzats $\Phi=e^{-i\omega t}e^{i k \phi}(\cosh y)^{-1/2} \psi
(y)$,
where $k=2\pi {\tt l}/L$ and $\tt l$  is any integer, and switching to
a new  radial coordinate
\be
z=\int {dy\over \sqrt{\sinh^2y+\lambda^2}}=F(\arcsin ({\sinh y\over
\sqrt{\sinh^2y+\lambda^2}}),\sqrt{1-\lambda^2})~~
\lb{19}
\ee
we find  that the radial function $\psi(z)$
should satisfy effective Schr\"odinger  equation
\be
\partial_z^2\psi(z)+(\omega^2-U(z))\psi(z)=0~~,
\lb{20}
\ee
where
\be
U(y)={3\over 4}\cosh^2y+k^2+{1\over
4}(\lambda^2-2)-(1-\lambda^2)({1\over
  4}+k^2){1\over \cosh^2y}
\lb{21}
\ee
is the effective radial potential. Since there is no horizon in the
metric
(\ref{9}) the quasi-normal boundary conditions are no more in place.
Instead
we demand that solution to the radial equation (\ref{20}) be
normalizable
which means it should fall off appropriately at infinity .

The integration in (\ref{19}) results in some elliptic function. For
our
purposes it is however convenient to use an approximation valid in the
case
when $y\ll 1$. Notice that even though $y$ is small it can  either be
as small
as $\lambda$, $y\sim \lambda$, or be much larger than $\lambda$,
$y\gg\lambda$;
in the second case we have $\lambda\ll y\ll 1$. So that regime of small
(compared to $1$) $y$ gives us possibility to probe both the throat
region
(set by $\lambda$) and the outside the throat region with the size set
by
$1$ in terms of the coordinate $y$. For $y\ll 1$ we can replace
$\sinh^2
y\simeq y^2$ in (\ref{19}). Then the integration in (\ref{19}) is
easily
performed
\be
z=\int{dy\over \sqrt{y^2+\lambda^2}}={\tt arcsinh}\ {y\over
  \lambda}~\longrightarrow~y=\lambda\ \sinh z~~.
\lb{22}
\ee
Notice that $z$ defined this way can be both small and large.

In this approximation the radial potential takes the form
\be
U(z)\simeq \lambda^2(k^2+{1\over
  2})+\lambda^2(k^2-k^2\lambda^2+1-{\lambda^2\over 4})\sinh^2 z~~.
\lb{23}
\ee
The Schr\"odinger equation (\ref{20}) with this potential can be
written in the
form
\be
\partial_z^2\psi(z)+(\tilde{\omega}^2-B^2\sinh^2z)\psi(z)=0
\lb{23'}
\ee
where we defined
$$
\tilde{\omega}^2=\omega^2-\lambda^2(k^2+{1\over
  2})~,~~B^2=\lambda^2(k^2-k^2\lambda^2+1-{\lambda^2\over 4})~~.
$$
We remind that in the absence of the throat the near-horizon potential
is
given by exponentially decaying function, $U(z)\sim e^{2z}$.
The Schr\"odinger equation with such potential does not have a discrete
spectrum.
One of the effects which the throat  has produced is to replace  this
exponential function by the potential $U(z)\sim \lambda^2 \sinh^2 z$
which has
the form of the potential well and thus admits the discrete spectrum.
Equation (\ref{23'})  describes the quantum mechanical Toda Lattice.
The discrete spectrum can be found explicitly using technique
developed\footnote{I
  thank M. Olshanetsky for pointing this reference to me.} in
\cite{Gutzwiller}. Rather than deal with  exact analysis we however
prefer
to apply the WKB prescription and get the spectrum approximately.
The approximation is accurate in the limit of small $\lambda$.
In what follows we assume that $k=0$ (${\tt l}=0$).

\subsection{The spectrum in the  WKB approximation}
For the equation (\ref{23'}) the WKB prescription gives us the
quantization
condition
\be
\int dz \sqrt{\tilde{\omega}^2-B^2\sinh^2z}=\pi (n+{1\over 2})~,~~n\in
{\bf
  Z}
\lb{w1}
\ee
where the integration is taken over $z$ for which the expression
staying
under the square root is positive. The integration can be performed
explicitly
in terms of elliptic functions,
\be
J(a)=\int_0^{{\tt arcsinh} (a)}\sqrt{a^2-\sinh^2z}dz=\sqrt{a^2+1}({\bf
K}({a\over
  \sqrt{a^2+1}})-{\bf E}({a\over \sqrt{a^2+1}}))~~,
\lb{w2}
\ee
where we introduced $a=\tilde{\omega}/B$. For higher energy levels
$\omega\gg\lambda $ we have
that $a\simeq{\omega\over \lambda }\gg 1$ and can use the asymptotic
formula
for elliptic functions to get asymptotic expression
$$
J(a)\simeq a\ln(4a)
$$
valid for large values of $a$. Assuming that $\omega \ll
 (1/\lambda) $ the WKB quantization condition (\ref{w1}) then produces
the
 spectrum
\be
\tilde{\omega}={\pi\over 2\ln{4\over\lambda}}(n+{1\over 2})~,~~n\in
{\bf Z}~~
\lb{w3}
\ee
in agreement with our qualitative analysis in section 3.
Notice that we could have used the WKB prescription for the whole
potential (\ref{21})
and then assumed that the frequency $\omega\ll 1$. This condition than
would bring
the essential region in the WKB integral close to the bottom of the
potential
where the potential can be approximated by (\ref{23}).

The frequency $\tilde{\omega}$ is not the same as the frequency
$\omega$ which appears in the radial
Schr\"odinger equation (\ref{20}), both are related as
$\omega^2=\tilde{\omega}^2+\lambda^2/2$. The spectrum for the frequency
$\omega$ then can be represented in the following suggestive form
\be
&&\omega^2_n=m^2+p_n^2\nonumber \\
&&m^2={\lambda^2\over 2}~,~~p_n={\pi\over
2\ln{4\over\lambda}}(n+{1\over
  2})~,~~n \in {\bf Z}~~.
\lb{w4}
\ee
The spectrum thus is that of massive particle.
By our assumption, (\ref{w4}) should be identified with spectrum
(understood
as poles in 2-point function) of the
boundary theory at finite $\tt k$ and is our prediction.
Note that the spectrum (\ref{w4}) relies only on the shape of the
modified
metric in the near-horizon region and in this sense is universal.
The parameter $\lambda({\tt k})$ appears both in the quantization of
the momentum $p_n$
in the $z$-direction and in the mass. The appearance of the mass $m\sim
\lambda$ is in
agreement with our analysis
in section 4. Notice that the inverse mass $1/m=t_P$ and the
inverse momentum $1/p_0=t_H$ determine two radically different time
scales\footnote{There is, of course, one more time scale in the game:
it is
  set by size $L$ ($\sim \sqrt{MG}$) of black hole. This time scale
appears
  due to $k\neq 0$ (${\tt l}\neq 0$)  part in spectrum of equation
(\ref{20}) and
is much smaller than $t_H$ and $t_P$.}:
$t_P\sim 1/\lambda\sim e^{{\tt k}L}$ and $t_H\sim \ln{1\over
\lambda}\sim {\tt k}L$, $t_P\gg t_H$. Comparing
both scales let's assume that $\lambda\sim 10^{-10}$ then the time
scale related
to the the inverse momentum, $t_H\sim 10$, while the time scale
determined by the mass, $t_H\sim 10^{10}$. Since the mass $m$ in
(\ref{w4})
is extremely small the spectrum determined by (\ref{w4}) describes
(almost) periodic
evolution with the period set by the time scale $t_H$. The mass $m$ is
however
non-vanishing and hence this evolution is not exactly periodic, the
ratio
of any two frequencies $\omega_n/\omega_k$ is not given by rational
number
in general, and hence the time evolution of the system is actually
quasi-periodic.
It is important that on the time scale much larger than $t_H$ the
evolution is
dominated by the periodicity with much larger period $t_P$. The latter
is the
longest period in the system and is thus
naturally associated with the Poincar\'e recurrence time.

\subsection{The mass and  large time scale periodicity}
In this subsection we want to illustrate this last statement, namely
that in the
case when $1/m\gg 1/p_0$ the large time scale evolution of the system
is
periodic with the period set by inverse mass $1/m$.
We will do this on the boundary theory side.
For simplicity we consider
massive scalar field on circle of size $L$ (note that in this
sub-section $L$ is similar to
what we earlier denoted as $L_{\tt eff}$) with periodic boundary
conditions.
In two dimensions the conformal dimension of scalar field is zero.
Therefore, the correlation function of two such fields has logarithmic
divergence. Correlation function of fields with higher conformal
dimension can
be obtained by differentiation of the correlation function of scalar
field.
For instance, the 2-point function of massive fermions (conformal
dimension $1/2$) is given by
\be
S_F(x,x')=(i\gamma^a\partial_a+m)G_F(x,x')~~.
\lb{m1}
\ee
The time periodicity of $G_F$ is preserved in $S_F$.
We choose the massive scalar field with periodic boundary condition
because the spectrum
in this case
\be
\omega^2_n=m^2+({2\pi n\over L})^2~,~~n\in {\bf Z}
\lb{m0}
\ee
is similar to the spectrum (\ref{w4}) ($L\sim \ln{1\over \lambda}$ and
$m\ll 1/L$).
The 2-point function (or, in the case at hand, the Feynman propagator)
of scalar field satisfying the periodic boundary
condition takes the form
\be
G_F(x,x')=-\sum_{n=-\infty}^{+\infty}{1\over
  4}H^{(2)}_0(m\sqrt{(t-t')^2-(\phi-\phi'+L n)^2-i\epsilon})~~
\lb{m2}
\ee
of sum over images to maintain the periodicity in $\phi$.
In momentum space the sum over $n$ appears as sum over all poles
(\ref{m0}).
For simplicity we ignore the temperature which otherwise should be
taken into
account by imposing the condition of periodicity in Euclidean time with
period
$1/T$. Let us set $\phi=\phi'$ and $t'=0$ then the correlation function
(\ref{m2})
is function of only time
\be
G_F(t)=-\sum_{n=-\infty}^{+\infty}{1\over
  4}H^{(2)}_0(m\sqrt{t^2-L^2 n^2-i\epsilon})~~.
\lb{m3}
\ee
Manipulating further with this expression we first go to Euclidean
time $t=i\tau$, the Hankel function $H^{(2)}_0(m\sqrt{t^2-L^2n^2})$
becomes  MacDonald function $K_0(m\sqrt{\tau^2+L^2n^2})$ under this
analytic continuation.
Then we  replace the infinite sum over $n$ by integral, this procedure
gives a good approximation if $\tau/L\gg 1$. Thus we have that
$$
\sum_{n=-\infty}^{+\infty}K_0(m\sqrt{\tau^2+L^2 n^2})\simeq
2\int_0^\infty dn K_0(mL\sqrt{{\tau\over
    L}^2+n^2})={\pi\over mL}e^{-m\tau}~~,
$$
where in the last passage the integral over $n$ was performed
explicitly using
formula (6.596.3) from \cite{GR}.
Analytically continuing back to the real
time we find that
\be
G_F(t)\simeq -{\pi\over 4mL}e^{-imt}~~.
\lb{m4}
\ee
This is the desired formula which describes large $t$ ($t\gg L$)
behavior of the
correlation function. Clearly this behavior is periodic with the period
set
by  inverse mass $1/m$. This periodic behavior is a result of
superposition
of contributions from large number of images in (\ref{m3}).

\subsection{Black Hole Poincar\'e recurrences}
In general, system with discrete frequency spectrum shows rather
complicated time
evolution. Being quasi-periodic in nature  it may  look dissipative on
certain
time scales. Example of this we have seen in section 2.2.2 for free
fermions
on circle. For  more complicated system  the characteristic time is the
so
called Heisenberg time (discussed in detail in \cite{Barbon})
which can be defined as $t_H=1/\<\delta\omega\>$ where
$\delta\omega_{kn}=\omega_n-\omega_k$ is the transition frequency and
some
sort of averaging over the spectrum is assumed. The Heisenberg time is
the time scale which
characterizes the discreteness of the spectrum. For time $t\ll t_H$ the
spectrum can be approximated as continuous. In particular, this means
that
for time $t\ll t_H$ the system may show dissipative behavior similar to
the
one we have observed in section 2.2.2.  For larger time $t\gg t_H$ the
intrinsic quasi-periodicity in the system becomes more visible and the
time
evolution of correlation functions starts to show long-period
oscillations. The longest one is
given by the Poincar\'e recurrence time $t_P$.

Returning to the spectrum (\ref{w4}) we see that it gives a
particularly
simple example of  time evolution we have just described.
To  the leading order the spacing between energy levels is given by
$\pi/\ln
(1/\lambda )^2$ so that the Heisenberg time $t_H$ is related  to the
time scale
$\ln (1/\lambda)$ we
defined in section 5.1.  The black hole then can be approximated  by
continuous
spectrum on the time scale $t\ll t_H=\ln{1\over \lambda}$ and thus
shows the usual
non-unitary (thermodynamic) behavior typical for space-times with
semiclassical
horizons.  In particular, it may be characterized by complex
quasi-normal
modes if observed during  time $t\ll t_H$.
Time set by size $L$ of black hole is the main characteristic time
scale in
this regime.
The discreteness of the spectrum becomes manifest on the
time scale close to $t_H$ so that the correlation functions start to
demonstrate certain periodicity. The spectrum looks as equidistant on
this
time scale. But even then unitarity is not
yet completely restored: there is more information hidden in much
longer oscillations.
These oscillations are due to the fine structure of the spectrum
(\ref{w4})
which deviates from  that of
equidistant: the non-vanishing mass in (\ref{w4}) drives the largest
time scale
correlations in the system. The time scale $t_P=1/m\sim e^{1\over 4 G}$
is thus the Poincar\'e
recurrence time during which all information available
in the system (black hole) is released.
This is  the time scale on which evolution of black hole is
{\it  ultimately } unitary. Note that the brick wall produces exactly
equidistant spectrum and thus does not give rise naturally to the
hierarchy
of time scales. In our picture the latter comes out as a result of the
non-semiclassical (smooth) modification (\ref{9}) of the near-horizon
geometry.

\section{Conclusions}

We conclude with several remarks.
As is well known the black hole in AdS space can be in thermal
equilibrium
with the Hawking radiation and thus represents a well-defined example
of what is
called eternal black hole. It is a great simplification to us since  in
the non-semiclassical modification of black hole geometry we may
restrict our
consideration to static case and do not
consider dynamical evolution of black hole due to quantum evaporation.
This evolution can be rather complicated and it is not yet clear
how the non-semiclassical modification should work in this case.
It is however an interesting problem which we are planning to analyze
in the future.

It is of course natural to ask how our consideration extends to other
spacetimes with horizons, for instance whether de Sitter space time
could be
understood along same lines. Cautiously, we might expect that our
picture may be
useful in this case as well although details may be more subtle and yet
have
to be worked out.

Semiclassical singularity at $r=0$ did not play any
role in our consideration. It is because, as we discussed this before,
region with $r^2\leq 1-\lambda^2$ is now disconnected from the region
$r^2\geq 1$ which is main focus  in this paper. On the other hand,
space time
(\ref{9}) does contain region of trans-Planckian curvature which should
manifest itself somehow. For instance, it may be useful to consider two
copies
of CFT living on two asymptotic boundaries ($y=-\infty$ and
$y=+\infty$).
Semiclassically they are separated by horizon but are able to
communicate
through semiclassical singularity, the correlation function between two
CFT
thus contains information about the singularity (see \cite{sing} for
more details).
In non-semiclassical spacetime (\ref{9}) the two theories can talk to
each
other directly. Geodesics connecting two boundaries  pass through the
highly
curved region so that the correlation functions between two theories
should carry information about the trans-Planckian curvature. It would
be
interesting to see how this information is encoded in the correlation
functions.

An interesting question is what happens to the semiclassical $SL(2,{\bf
Z})$ symmetry
and does it make sense to consider the T-dual of the metric (\ref{9}).
This boils down to clarifying the non-perturbative status of the
$SL(2,{\bf Z})$ symmetry.
Semiclassically the T-dual to BTZ metric is  thermal AdS. The formal
T-dual of the metric (\ref{9}) is some deformation of AdS space-time
with main deformation concentrated near the origin. Since the origin in
AdS
is not in any sense a special point this modification might be
difficult to
justify. One possibility can be that the deformation (\ref{9}) of BTZ
metric
is a manifestation of particular choice of the quantum state of black
hole,
this choice of state may not be
natural in the case of AdS spacetime. The symmetry then should
refer to spacetimes with  same choice of the quantum state.

Important open question is whether the metric (\ref{9}) can be
consistently justified within
string theory. Whatever the answer is we believe that the findings
in this paper provide us with sort of existence theorem to the solution
of the unitarity problem.

\bigskip

\bigskip

\noindent{\bf \Large Acknowledgments}

\bigskip

\noindent I would like to thank J. de Boer, N. Kaloper, K. Krasnov and
I. Sachs for inspiring discussions and helpful remarks. I thank  E.
Donets
and M. Olshanetsky for help with  the Toda
Lattice and its spectrum. A email correspondence with M. Porrati is
greatly acknowledged.

\end{document}